\documentclass[prb]{revtex4}
\def\no{\noindent}
\def\bc{\begin{center}}
\def\ec{\end{center}}

\def\beq{\begin{equation}}
\def\eeq{\end{equation}}

\begin{document}
\title{
Two-component Bose gas in an optical lattice at single-particle filling}

\author{K. Ziegler}
%\author{Second Author}
\affiliation{Institut f\"ur Physik, Universit\"at Augsburg, Germany}

%\date{\today}
\begin{abstract}
The Bose-Hubbard model of a two-fold degenerate Bose gas is studied
in an optical lattice with one particle per site and virtual tunneling to
empty and doubly-occupied sites. An effective Hamiltonian for this system 
is derived within a continued-fraction approach.
The ground state of the effective model is studied in mean-field approximation
for a modulated optical lattice. A dimerized mean-field state gives
a Mott insulator whereas the lattice without modulations develops
long-range correlated phase fluctuations due to a Goldstone mode. 
This result is discussed in comparison with the superfluid and the 
Mott-insulating state of a single-component hard-core Bose.
\end{abstract}
\pacs{03.75.Mn, 05.30.Jp}

\maketitle

\section{Introduction}

\no
An ultracold Bose gas is brought into an optical lattice, created by 
a stationary Laser field \cite{greiner02}. If the corresponding 
periodic potential is sufficiently strong, such that tunneling of atoms 
between the potential wells is strongly suppressed, the phase coherence of 
the Bose gas is destroyed due to the repulsive interaction between the bosons.
In this case the Bose gas becomes a Mott insulator. This
ground state is characterized by a fixed number of bosons in each lattice
well and strong incoherent phase fluctuations of the quantum state,
in contrast to the phase coherence of the Bose-Einstein condensate 
with a strongly fluctuating local particle number.
The Mott insulator and the transition to the superfluid state were discussed
theoretically some time ago \cite{fisher89,ziegler93,jaksch98} and observed 
experimentally recently \cite{greiner02}.

Besides the phase of the bosons and the local particle
number there can be other degrees of freedom in an ultracold gas of bosonic
atoms which may also establish some long-range ordering. A possible 
candidate for such a consideration
is a Bose gas in an optical or a magnetic trap \cite{ketterle00},
where the fluctuations between nearly degenerate hyperfine states represent
an additional degree of freedom. This can play a role in establishing new 
types of ordering, similar to the spin degree of freedom in fermionic
systems. Here the case of a strongly interacting Bose gas with one
particle per site will be considered. Using a Bose-Hubbard model,
the interaction and the chemical potential of a grand canonical ensemble
are ajusted such that there is
one particle per optical lattice site. According to the statements
given above, the fixed number of particles per site would represent a 
Mott insulator. On the other hand, the local particle number $n_{\bf r}$
is a sum of the particle numbers of both components (represented by
a ''spin'' $\uparrow$ or $\downarrow$)
\[
n_{\bf r}=n_{{\bf r},\uparrow}+n_{{\bf r},\downarrow}.
\]
The individual particle numbers of the two components are fluctuating 
quantities and can lead to a new state when long-range
correlations develop. A similar situation can be found in a single-component
hard-core Bose gas in an optical lattice. Then each lattice site is either
empty with the quantum state $|0\rangle$ or singly occupied with $|1\rangle$. 
Formally these two states correspond with the states $|\uparrow\rangle$ and
$|\downarrow\rangle$ of the two-component gas. It is known from analytic
\cite{fisher89,ziegler93} and numerical calculations \cite{schmid02} that
the hard-core Bose gas develops a superfluid phase for abitrarily small
tunneling rates if the states $|0\rangle$ and $|1\rangle$ are degenerate, 
i.e. when $\langle n_{\bf r}\rangle = 1/2$. This behavior will be discussed in 
Sect. 3.1. Using the formal correspondence with the two-component Bose gas the
development of a long-range correlated state would also take place for any 
tunneling 
rate if the states $|\uparrow\rangle$ and $|\downarrow\rangle$ are degenerate.
A consequence would be that these two states could easily separate in
space, leading to an entangled state in the Bose gas.

Ordering phenomena can be studied perturbatively, starting from isolated
potential wells of the optical lattice and systematically turning on the 
tunneling between these potential wells. Because of the degeneracy with
respect to the spin degrees of freedom in the isolated wells this requires 
a degenerate perturbation theory. Instead of using a perturbation theory for 
the tunneling Hamiltonian a continued-fraction approach will be applied
in the following.

The central aim of this paper is to derive an effective Hamiltonian for the
two-component Bose gas with $\langle n_{\bf r}\rangle=1$ and to discuss the 
properties of this system. In this case a particle is a superposition of the 
states $|\uparrow\rangle$ and $|\downarrow\rangle$. Without tunneling between 
lattice sites it represents a ''paramagnetic'' state, i.e. a state without 
ordering. Tunneling, on the other hand, preserves the ''spin'': starting 
with a boson of a given spin this boson will spread to neighboring potential 
wells. This process requires virtual states which are empty or occupied by
more than one particle. Here only occupation with two particles will be
allowed to keep the calculational effort low. However, the continued-fraction 
approach proposed in this paper can be extended to higher orders
of occupation.

After introducing the two-component Bose-Hubbard model in Sect. 2 the 
single-component hard-core Bose gas is considered for comparison in Sect. 3. 
and treated in mean-field approximation in Sect. 3.1. The ground states of 
a two-component Bose-Hubbard model with $\langle n_{\bf r}\rangle\le 1$ 
and a single-component hard-core Bose gas are evaluated for a two-site system 
in Sect. 3.2. Then the effective Hamiltonian for a projected system of 
interacting particles is derived by a truncated continued fraction in
Sect. 4 and applied to the two-component Bose gas with 
$\langle n_{\bf r}\rangle=1$ in Sect. 5. Finally, the two-site system 
(Sect. 5.1) and a mean-field approximation (Sect. 5.2) are 
discussed for the effective model.

\section{The Model: Interacting Bose Gas}

In order to describe a multi-component interacting Bose gas 
the interaction of bosons can either be included as a hard-core
interaction or in terms of the Bose-Hubbard model. Crucial is only that
a repulsive interaction stabilizes an incompletely filled lattice with 
vanishing compressibility. From this point of view the actual form of $H_0$ is
not important except for the fact that it depends only on the local particle
number at site ${\bf r}$
\[
n_{\bf r}=\sum_{\sigma=\downarrow,\uparrow} 
a^\dagger_{{\bf r},\sigma} a_{{\bf r},\sigma}
\]
with boson creation (annihilation) operators $a^\dagger$ ($a$). For a 
more specific discussion, a Bose-Hubbard model with
\beq
H_0= \sum_{\bf r} [-\mu n_{\bf r}+Un_{\bf r}(n_{\bf r}-1)]
\label{ham0}
\eeq
shall be considered, where $\mu$ is the chemical potential and $U>0$ 
the interaction constant. Eigenvalues of the local particle number
are $n=0,1,...$ with corresponding energies per lattice site 
\[
E(n)=-\mu n+Un(n-1).
\]
For $0<\mu<U$, the case considered throughout this paper, the lowest energy 
is $E(1)=-\mu$ for $n=1$, and next higher energies are $E(0)=0$ and
$E(2)=2(U-\mu)$ and even higher energies for $n>2$. 

The dynamics of the bosons is described by the tunneling Hamiltonian
\beq
H_1=-\sum_{<{\bf r},{\bf r}'>}\tau_{{\bf r},{\bf r}'}
\sum_{\sigma=\downarrow,\uparrow} t_\sigma
a^\dagger_{{\bf r},\sigma} a_{{\bf r}',\sigma},
\label{ham1}
\eeq
where $<{\bf r},{\bf r}'>$ are nearest-neighbor sites on the optical lattice.
The parameter $0\le\tau_{{\bf r},{\bf r}'}\le1$ describes a modulation of the
optical lattice. The 
interaction is crucial for finding new physical states, since the
noninteracting Bose gas (i.e. for $U=0$) the Hamiltonian $H_1$
gives only two independent Bose-Einstein condensates.

A grand-canonical ensemble of bosons at the inverse temperature
$\beta$, defined by the partition function 
\beq
Z=Tr \exp(-\beta H),
\label{partition}
\eeq
can be used to evaluate the average density of particles as
\[
{\bar n}={1\over \beta N}{\partial log Z\over \partial\mu}
\]
with the number of lattice site $N$. Without tunneling (i.e. for $t_\sigma=0$) 
the average density of particles ${\bar n}$ 
gives $\partial {\bar n}/\partial\mu=0$ (i.e. incompressible states)
for all non-integer values of $\mu/U$.

\section{Single-component Hard-Core Bose Gas: Superfluid and 
Mott-insulating States}

Before discussing the two-component Bose gas the single-component hard-core
Bose gas shall be considered because its properties are known from a
number of other approaches.
The hard-core interaction can be described by a bosonic creation (annihilation)
operator $A^\dagger$ ($A$) with the additional condition ${A^\dagger}^2=0$. 
The Hamiltonian of the hard-core Bose gas then reads
\beq
H_{hcb}=-t\sum_{<{\bf r},{\bf r}'>}A^\dagger_{\bf r} A_{{\bf r}'}
-\mu\sum_{\bf r} A^\dagger_{\bf r} A_{\bf r}
\label{xyrep}
\eeq
and acts
on the Hilbert space with $|0\rangle$ and $|1\rangle$ as basis states at
each lattice site. There is a close formal connection between hard-core Bose 
and spin-1/2 operators, since one can write
\beq
S^x=(A+A^\dagger)/2,\ \ S^y=i(A-A^\dagger)/2,\ \ S^z=A^\dagger A-1/2.
\label{spin}
\eeq
The Hamiltonian can be expressed in terms of these spin operators as 
an XY model with a magnetic field in $z$ direction:
\[
H_{hcb}=-t\sum_{<{\bf r},{\bf r}'>}(S^x_{\bf r}S^x_{{\bf r}'}
+S^y_{\bf r}S^y_{{\bf r}'})-\mu\sum_{\bf r} S^z_{\bf r}.
\]
For sufficiently large values of $|\mu|$ the ground state is ferromagnetic:
if $\mu>0$ the magnetization is $\langle S^z\rangle>0$ and vice versa.
Therefore, for $\mu>0$ ($\mu<0$) these ground states correspond in terms
of the hard-core bosons with a completely filled (empty) lattice, where the
filled lattice represents a Mott-insulating state \cite{fisher89,ziegler93,
jaksch98}.
$\mu=0$ is a marginal situation, where the bosons develop a superfluid state
for any positive value of the tunneling rate $t$.
More general, the superfluid state persists if the tunneling dominates. 

The advantage of the spin representation is that it provides a simple
qualitative
picture for the existence of a Mott-insulating state and a transition
to a superfluid state. A more quantitive description is obtain from a 
mean-field approximation which is discussed in the next section.

\subsection{Mean-field approximation of the hard-core Bose gas}

A possible complex mean-field state for the hard-core Bose gas is
\beq
|\Psi_{MF}\rangle=\prod_{\bf r}\Big[e^{i\varphi_{\bf r}}\cos(\eta_{\bf r})
+e^{i\psi_{\bf r}}\sin(\eta_{\bf r})A_{\bf r}^\dagger\Big]|0\rangle
\label{mfstate}
\eeq
from which matrix elements can be calculated. For instance, the tunneling
term in the Hamiltonian $H_{hcb}$ gives
\[
\langle\Psi_{MF}|A_{\bf r}^\dagger A_{{\bf r}'}+A_{{\bf r}'}^\dagger A_{\bf r}
|\Psi_{MF}\rangle
=2\cos(\alpha_{\bf r}-\alpha_{{\bf r}'})\cos(\eta_{\bf r})\sin(\eta_{\bf r})
\cos(\eta_{{\bf r}'})\sin(\eta_{{\bf r}'}),
\]
where the phases appear only in the phase difference 
$\alpha_{\bf r}=\varphi_{\bf r}-\psi_{\bf r}$. Thus this term of the 
Hamiltonian 
has a global $U(1)$ symmetry because it is invariant under a shift
$\varphi_{\bf r}\to\varphi_{\bf r}+\Delta$ and
$\psi_{\bf r}\to\psi_{\bf r}+\Delta'$.
The other matrix element of the Hamiltonian $H_{hcb}$ is independent 
of the phases:
\[
\langle\Psi_{MF}|A_{\bf r}^\dagger A_{\bf r}|\Psi_{MF}\rangle
=\sin^2\eta_{\bf r}.
\]
The mean-field expectation of the Hamiltonian (\ref{xyrep}) with the
homogeneous mean field $\eta$ reads
\[
\langle\Psi_{MF}| H_{hcb}|\Psi_{MF}\rangle
=-\sin^2\eta\Big[
t(1-\sin^2\eta)\sum_{<{\bf r},{\bf r}'>}\cos(\alpha_{\bf r}-\alpha_{{\bf r}'})
+\sum_{\bf r}\mu
\Big].
\]
The ground state is $\cos(\alpha_{\bf r}-\alpha_{{\bf r}'})=1$ and
\beq
\sin^2\eta =\cases{
0 & for $\mu\le -2dt$ \cr
1/2+\mu/4dt & for $-2dt<\mu<2dt$ \cr
1 & for $2dt\le\mu$ \cr
}.
\label{hcphase}
\eeq
The expression
\beq
\langle\Psi_{MF}|A_{\bf r}|\Psi_{MF}\rangle
=e^{i(\psi_{\bf r}-\varphi_{\bf r})}\cos \eta\sin\eta
\label{op}
\eeq
is an order parameter for a superfluid state which vanishes in the cases
$\sin^2\eta =0,1$. It should be noticed that in this regime 
$\langle\Psi_{MF}| H_{hcb}|\Psi_{MF}\rangle$
is independent of 
$\alpha_{\bf r}$. This reflects the absence of a superfluid.
In the case $\sin^2\eta =0$ it is an empty state
and for $\sin^2\eta =1$ a Mott insulator with one particle per site. 
This result is in agreement
with Monte Carlo simulations \cite{schmid02}.

\subsection{Hard-core vs. Bose-Hubbard model: a two-site system}

To study the tunneling between neighboring potential wells of the optical
lattice a two site model with $t_\uparrow=t_\downarrow\equiv t$
is considered. This was already discussed in great 
detail for the model under consideration, using the leading order of an 
expansion in $t^2/U$ \cite{duan02,kuklov03,altman03}. Here the study 
shall be performed without
assuming $t^2/U\ll 1$. As a first example the single-component hard-core
Bose gas is considered in terms of the Hamiltonian in Eq. (\ref{xyrep}). 
The ground states are
\[
|\Psi_0\rangle=\cases{
|0,0\rangle & for $\mu< -t$ \cr
(|0,1\rangle +|1,0\rangle)/\sqrt{2} & for $-t<\mu <t$ \cr
|1,1\rangle & for $\mu>t$
}.
\]
%whereas it is the empty state $|0,0\rangle$ for $\mu<-t$ and the
%fully occupied state $|1,1\rangle$ for $\mu>t$. 
This result corresponds with the three different mean-field ground states 
in (\ref{hcphase}). Thus already the two-site system indicates the
three phases of the full $d$-dimensional lattice: the empty lattice,
the condensate and the $n=1$ Mott insulator. 
An analogous calculation for the two-component  
Bose-Hubbard model with empty and singly-occupied sites gives the
ground states
\beq
|\Psi_0\rangle=\cases{
|0,0\rangle & for $\mu<-t$ \cr
(|0,\uparrow\rangle+|\uparrow,0\rangle)/\sqrt{2},\ \ 
(|0,\downarrow\rangle+|\downarrow,0\rangle)/\sqrt{2} 
& for $-t<\mu<t$ \cr
|\downarrow,\downarrow\rangle, \ \ |\downarrow,\uparrow\rangle, \ \ 
|\uparrow,\downarrow\rangle, \ \ |\uparrow,\uparrow\rangle
& for $\mu>t$ \cr
}.
\label{bhphase}
\eeq
For $-t<\mu<t$ there are two (degenerate) ferromagnetic states, where 
the degeneracy can be lifted by an infinitesimal magnetic field. 
It is expected that these are the states which form a condensate.
The four-fold degeneracy for $\mu>t$
may be lifted when a virtual tunneling through empty and doubly-occupied 
sites is included. This degeneracy also raises the question whether or not an
analogue of the
superfluid ground state is allowed due to virtual tunneling in this regime.
A reason for having long-range correlations is that one of the two components, 
e.g. $|\downarrow\rangle$, can be formally considered as an empty site,
the other component as a hard-core boson and since the hard-core Bose gas 
has a superfluid state for sufficiently large tunneling rate.
To study this regime, a projection of the trace in the partition function to 
singly-occupied states is considered subsequently. This projection allows 
virtual tunneling through empty and doubly-occupied sites.

\section{A Continued-Fraction Approach to the Projected Partition Function}

The many-body system is defined by the Hamiltonian $H$ and the
transfer matrix $e^{-H}$. Physical quantities at inverse temperature
$\beta$ are derived from the partition function $Z$ defined in Eq. 
(\ref{partition}). A continued-fraction approach shall be developed in this
section to derive an effective Hamiltonian from $H$ that describes the
physics of a projected transfer matrix $P_0e^{-\beta H}P_0$.
Although $H$ is bounded from below it may have negative eigenvalues.
A positive operator can be obtained by adding a constant diagonal term 
$E$ to shift the ground state energy $E_0$ to positive values $E(1)=E_0+E$. 
Then the transfer matrix can be represented by the integral
\[
e^{-\beta H}={e^{\beta E}\over 2\pi i}\int_{-\infty}^\infty 
e^{i\beta z}(z-i(H+E))^{-1}dz.
\]
Thermodynamic properties at low temperatures are dominated by the ground
states and low-energy excitations. The trace of the grand-canonical
partition function $Z$ includes states with all possible number of
bosons. For the Hamiltonian $H$ the highest statistical weight comes
from the ground state. If $H=H_0+H_1$ with a perturbation $H_1$,
the Hilbert space is projected on the degenerate ground states of the 
Hamiltonian $H_0$ by $P_0$. In the example of Sect. 2 the
Hamiltonian preserves the number of particles. Therefore, in this case it is
expected that the
system with fixed particle filling gives the dominant contribution to the 
trace, especially at low temperatures. This situation can be described by 
the $P_0$-projected partition function with the corresponding trace $Tr_0$
\beq
Z=Tr_0(P_0 e^{-\beta H} P_0)
={e^{\beta E}\over 2\pi i}\int_{-\infty}^\infty 
e^{i\beta z}Tr_0[P_0(z-i(H+E))^{-1}P_0]dz.
\label{int2}
\eeq
Assuming that $H$ is implicitely shifted by $E$,
the $P_0$-projection of the resolvent $(z-iH)^{-1}$ reads
\beq
P_0(z-iH)^{-1}P_0=
(P_0(z-iH)P_0 + P_0H P_1 (z-iH)_1^{-1}
P_1 HP_0)_0^{-1}
\label{pidentity}
\eeq
with $P_1={\bf 1}-P_0$. $(...)_{0,1}^{-1}$ is the inverse with respect to the 
$P_{0,1}$-projected space. This identity can be directly shown by 
a multiplication of the matrix and its inverse. It can be generalized to
a recurrence relation (s. Eq. (\ref{recurr}) of Appendix A) if the 
Hamiltonian $H$ satisfies the special conditions (\ref{recurse}). 
For the two-component Bose gas this is indeed the
case, since the Hamiltonian $H=H_0+H_1$ of Sect. 2 is of the special form
\beq
H=\pmatrix{
P_0H_0P_0 & P_0H_1 P_1 \cr
P_1 H_1P_0 & P_1 H P_1 \cr
},
\label{structure}
\eeq
where $P_0$ is the projection to the degenerate ground state of $H_0$ 
with $0<\mu<U$, i.e. one particle per site.
Thus the Hamiltonian $H_1$ is responsible for an interaction between the
$P_0$- and the $P_1$-projected Hilbert spaces. $H_0$, on the other hand, 
acts only inside the $P_0$-projected space. Starting from singly-occupied 
states,
the tunneling Hamiltonian $H_1$ can only create a pair of an empty and a
doubly-occupied site (PEDS). If $P_2$ is the projection from $P_1$ to one 
with only a single PEDS, the second term in the inverse matrix of Eq.
(\ref{pidentity}) reads
\beq
P_0H_1 P_1 (z-iH)_1^{-1} P_1 H_1P_0
=P_0H_1 P_2 (z-iH)_1^{-1} P_2 H_1P_0.
\label{pident3}
\eeq
In general, the operator $P_{2k+1}H_1P_{2k}$ creates a new PEDS in the 
Hilbert space with $k$ PEDSs. Thus the continued-fraction representation
of Appendix A, applied to the two-component Bose gas at single-particle
filling, is based on the creation of PEDSs.

In order to truncate the continued fraction the creation of new PEDS and
multiply-occupied sites is excluded. This is related to the approximation 
\[
P_2 (z-iH)_1^{-1} P_2\approx P_2 (z-iH_0)_1^{-1} P_2
={1\over z-i[(N-2)E(1)+E(0)+E(2)]}P_2,
\]
where $E(0)$ and $E(2)$ are the energies of $H_0+E$ for 
the empty and doubly occupied sites of Sect. 2. With Eqs. (\ref{pidentity})
and (\ref{pident3}) this gives for the $P_0$-projected resolvent of the
partition function 
\[
P_0(z-iH)^{-1}P_0\approx
(P_0(z-iH_0)P_0 + P_0H_1P_2[z-iH_0]^{-1}P_2H_1P_0)_0^{-1}
\]
\[
=(z-iE_0 + P_0H_1^2P_0/[z-i[(N-2)E(1)+E(0)+E(2)]])_0^{-1}.
\]
Here it has been used that
\[
P_0H_1 P_2 H_1P_0=P_0H_1^2P_0
\]
which follows from the fact that $H_1$ is off-diagonal with
respect to the $P_0$- and $P_2$-projected Hilbert spaces.
Then the $P_0$-projected partition function reads for low
temperatures (i.e. $\beta\sim\infty$) (s. Appendix B)
\beq
Z\sim {1\over 2}e^{-\beta (NE_0+\Delta E)}
Tr_0\Big[e^{-\beta H_{eff}}(1-\Delta EH_{eff}^{-1})
\Big]
\label{proz}
\eeq
with the effective Hamiltonian
\[
H_{eff}=-((\Delta E)^2+P_0H_1^2P_0)^{1/2}
=-\Delta E(1+P_0H_1^2P_0/\Delta E)^{1/2}
\]
and
\[
\Delta E=[E(0)+E(2)]/2-E(1).
\]
Since $P_0H_1^2P_0$ is a non-negative operator,
this result implies that the ground state of $-P_0H_1^2P_0$ is the ground 
state of $H_{eff}$.

If $P_0H_1^2P_0$ is small in comparison with $(\Delta E)^2$, a perturbation 
theory
with respect to the $P_0H_1^2P_0$ can be applied to $H_{eff}$. This leads
to an expansion with respect to $t^2/U$. 
In leading order the approximation is
\[ 
H_{eff}\approx -\Delta E-P_0H_1^2P_0/(2\Delta E),
\]
in agreement with the results of Refs. \cite{duan02,kuklov03,altman03}.

This method of deriving an effective Hamiltonian by projecting the
partition function is quite general as long as the Hamiltonian $H$ has
the structure shown in Eq. (\ref{structure}).
The specific case of a two-component Bose gas will be discussed subsequently.

%%%%%% TWO-COMPONENT BOSE GAS %%%%%%%%%%%%%

\section{The Effective Hamiltonian of the Projected Two-component Bose Gas}

Numerous possibilities were discussed in the literature for the creation 
of a two-component system in atomic gases 
\cite{matthews99,search01,mandel03,nikuni03}. 
For instance, if $^{87}$Rb is coupled to a radiation field there are pairs
of nearly degenerate hyperfine $|F,m_F\rangle$ states, namely 
$|\uparrow\rangle =|1,-1\rangle$ and $|\downarrow\rangle =|2,1\rangle$ 
\cite{matthews99,mcguirk03}
or $|\uparrow\rangle =|1,-1\rangle$ and $|\downarrow\rangle =|2,-2\rangle$
\cite{mandel03}.
$|\downarrow\rangle$ and $|\uparrow\rangle$ are formal notations to
specify the two (almost) degenerate states.
The interaction of the atoms does not depend on the states but only 
on the local density of the bosons. Therefore, the interaction in $H_0$ 
of Eq. (\ref{ham0}) is a good description.

In a dilute regime (i.e., for a filling less than one particle per site)
the interaction is weak. This opens the opportunity to apply a classical 
approach for the two-component condensate 
order parameter, leading to a lattice version of the Gross-Pitaevskii 
equation. The optical lattice means that the kinetic term has a special 
band dispersion 
$\epsilon({\bf k})$ in Fourier space, depending on the lattice, instead of the 
$k^2$ dispersion in the case without an optical lattice.
It is believed that this model has a ferromagnetic ground state with respect to
the two-component bosons \cite{stoof00}, a result that is also supported
by the result of the two-site system in Eq. (\ref{bhphase}). On the other 
hand, it is well known from the theory of the
fermionic Hubbard model that the type of spin order depends crucially
on the filling of the lattice \cite{fulde,gebhard}, and can change, for 
instance, from ferromagnetic to antiferromagnetic order by changing
the filling. In particular, the fermionic Hubbard model has an 
antiferromagnetic ground state at half-filling.
To study this effect in the two-component Bose gas in an optical 
lattice, the strongly interacting case with single-particle filling of 
the lattice shall be considered here. Then the two-component 
degeneracy has to be taken fully into account and the projection 
approach of the previous Sect. for the tunneling term $H_1$ should be applied. 
In this case the Hamiltonian $-P_0H_1^2P_0$ reads
\beq
-{1\over 4}\sum_{<{\bf r},{\bf r}'>}\tau_{{\bf r},{\bf r}'}^2
\sum_{\sigma,\sigma'=\downarrow,\uparrow}t_\sigma t_{\sigma'} P_0 
(a^\dagger_{{\bf r},\sigma} a_{{\bf r}',\sigma}
+a^\dagger_{{\bf r}',\sigma} a_{{\bf r},\sigma})
(a^\dagger_{{\bf r},\sigma'} a_{{\bf r}',\sigma'}
+a^\dagger_{{\bf r}',\sigma'} a_{{\bf r},\sigma'})
P_0
\label{pham}
\eeq
and $\Delta E=U$.
On the $P_0$-projected Hilbert space (i.e. the space with exactly one particle
per site) the operators
\[
A^\dagger_{{\bf r}}=P_0a^\dagger_{{\bf r},\uparrow}a_{{\bf r},
\downarrow}P_0,\ \ 
A_{{\bf r}}=P_0a^\dagger_{{\bf r},\downarrow}a_{{\bf r},\uparrow}P_0
\]
are creation and annihilation operators of hard-core bosons,
when $|\downarrow\rangle$ is formally identified with a vacuum
state $|0\rangle$ and $|\uparrow\rangle$ with a one-particle state
$|1\rangle$. As shown in
Appendix C, the operator $-P_0H_1^2P_0$ reads in terms of the hard-core 
Bose operators as
\beq
-P_0H_1^2P_0
=-\sum_{<{\bf r},{\bf r}'>}\tau_{{\bf r},{\bf r}'}^2\Big[
t_\uparrow t_\downarrow A^\dagger_{{\bf r}}A_{{\bf r}'}
+{t_\uparrow^2+t_\downarrow^2\over2}({\bf 1}-A^\dagger_{{\bf r}}
A_{{\bf r}})A^\dagger_{{\bf r}'}A_{{\bf r}'}
\Big].
\label{hcham}
\eeq
This Hamiltonian describes a hard-core Bose gas with a repulsive
nearest-neighbor interaction. The competition between tunneling (favors
particles) and nearest-neighbor repulsion (favors a ground state with
checkerboard order) leads to a complex situation. Similar to the 
single-component hard-core Bose gas, this can also be discussed in terms
of spin-1/2 states. The representation of 
Hamiltonian (\ref{hcham}) as a spin Hamiltonian via Eq. (\ref{spin}) 
gives an anisotropic Heisenberg Hamiltonian. 

\subsection{Projected two-site model}

The situation of the projected partition function of the two-site model
can be discussed and compared with the previous results for the hard-core
Bose gas and the two-component Bose-Hubbard gas. Using the 
hard-core Bose Hamiltonian of the projected model in Eq. (\ref{hcham}) 
with $t_\uparrow=t_\downarrow\equiv t$ and $\tau_{{\bf r},{\bf r}'}=1$, 
the corresponding $4\times 4$ matrix for the four differents states with 
$\downarrow$ and $\uparrow$ at the two sites has the eigenvalues 
$\{-2t^2,0,0,0\}$. 
The unique ground state of $-P_0H_1^2P_0$ with energy $-2t^2$ is
\beq
|\Psi_0\rangle=
{1\over \sqrt{2}}(|\downarrow,\uparrow\rangle +|\uparrow,\downarrow\rangle)
%\ \ {\rm for}\ \ \mu>t
.
\label{gstate}
\eeq
The $P_0$-projection has apparently selected a non-degenerate ground state
from the four degenerate ground states of the Bose-Hubbard model with one 
particle per site in Eq. (\ref{bhphase}). This is a consequence of the virtual
tunneling to empty and doubly-occupied states in the model with the projected 
partition function, which was not included in the derivation of the
state $|\Psi_0\rangle$ of
Eq. (\ref{bhphase}).
This state is not an eigenstate to $S^z_{\bf r}$ but has a vanishing
expectation for $S^z_{\bf r}$. This reflects the fact that the ground state
has no tendency to develop a ferromagnetic order.

The projected partition function (\ref{proz}) reads for this two-site
model
\[
Z\sim e^{-\beta(2E_0+U)}\sum_{j=1}^4 e^{\beta (U^2-E_j)^{1/2}}
\sim e^{-\beta(2E_0+U)}e^{\beta (U^2+2t^2)^{1/2}}.
\]
$Z$ can be used to evaluate the average tunneling energy from
\[
{t\over\beta}{\partial\over\partial t}\ln Z\sim{2t^2\over\sqrt{U^2+2t^2}}. 
\]
Thus the interaction reduces the tunneling rate of the two-site model by a 
factor $(1+U^2/2t^2)^{-1/2}$.

\subsection{Mean-field approximation of the two-component Bose gas}

Using the hard-core Bose representation of the Hamiltonian in Eq.
(\ref{hcham}) its mean-field approximation is studied with the complex 
mean-field state (\ref{mfstate}). The repulsive nearest-neighbor interaction is
\[
\langle\Psi_{MF}|({\bf 1}-A_{\bf r}^\dagger A_{\bf r})
A_{{\bf r}'}^\dagger A_{{\bf r}'}|\Psi_{MF}\rangle
=\cos^2\eta_{\bf r}\sin^2\eta_{{\bf r}'}.
\]
Together with the hard-core Bose Hamiltonian of Sect. 3.1 the expression in
the Hamiltonian of the two-component Bose gas at single-filling
reads in mean-field approximation
\[
-\langle \Psi_{MF}|P_0H_1^2P_0|\Psi_{MF}\rangle=
\]
\[
-\sum_{<{\bf r},{\bf r}'>}\tau_{{\bf r},{\bf r}'}^2\Big(
t_\uparrow t_\downarrow\cos(\alpha_{\bf r}-\alpha_{{\bf r}'})
\cos\eta_{\bf r}\sin\eta_{{\bf r}}\cos\eta_{{\bf r}'}\sin\eta_{{\bf r}'}
+{t_\uparrow^2+t_\downarrow^2\over 2}
\cos^2\eta_{\bf r}\sin^2\eta_{{\bf r}'}\Big).
\]
The first term favors a homogeneous solution for $\eta_{\bf r}$, 
the second term an inhomogeneous solution,
e.g., for neighboring sites ${\bf r},{\bf r}'$ with 
\beq
\sin^2\eta_{\bf r}=1,\ \ \cos^2\eta_{{\bf r}'}=1.
\label{sol2}
\eeq
In this case the first term vanishes and the remaining Hamiltonian is
\[
-\langle \Psi_{MF}|P_0H_1^2P_0|\Psi_{MF}\rangle_i=
-{t_\uparrow^2+t_\downarrow^2\over 4}
\sum_{<{\bf r},{\bf r}'>}\tau_{{\bf r},{\bf r}'}^2.
\]
On the other hand, a homogeneous mean-field solution 
for the ground state is $\sin^2\eta=1/2$ such that
\beq
-\langle \Psi_{MF}|P_0H_1^2P_0|\Psi_{MF}\rangle_h=
-{1\over4}\sum_{<{\bf r},{\bf r}'>}\tau_{{\bf r},{\bf r}'}^2\Big(
t_\uparrow t_\downarrow\cos(\alpha_{\bf r}-\alpha_{{\bf r}'})
+{t_\uparrow^2+t_\downarrow^2\over 2}\Big).
\label{ham2}
\eeq
This Hamiltonian
agrees formally with the mean-field Hamiltonian of hard-core bosons in 
Sect. 3.1 at the point of degeneracy $\mu=0$. However,
its interpretation in terms of the physical bosons, given by the Bose
operators $a^\dagger$ and $a$, is different. This is clearly indicated by
the fact that the order parameter of a superfluid state vanishes:
\[
\langle\Psi_{MF}|a_{\bf r}|\Psi_{MF}\rangle_h=0.
\]
Thus the long-range correlated phase fluctuations are not related to a 
superfluid state but
to a spontaneously broken symmetry, associated with the order parameter
\[
\langle\Psi_{MF}|A_{\bf r}|\Psi_{MF}\rangle_h
=\langle\Psi_{MF}|
a^\dagger_{{\bf r},\downarrow}a_{{\bf r},\uparrow}|\Psi_{MF}\rangle_h.
\]
%as given in Eq. (\ref{op}).
These phase fluctuations prevent the system to become a genuine
Mott insulator, since the latter is characterized by a gap and 
short-range correlated fluctuations (cf. with the Mott-insulating state
of the single-component hard-core gas in Sect. 3.1). However,
a Mott insulator can be obtained in the limit $t_\uparrow t_\downarrow=0$.
This is a limit similar to the Falicov-Kimball limit of the fermionic 
Hubbard model \cite{gebhard}.

Another mean-field approximation can be constructed for a generalization
of the two-site model to a modulated lattice model, using dimers with
tunneling rates $\tau_{{\bf r},{\bf r}'}=\tau_0$ as building blocks of 
the lattice. These dimers are weakly coupled with tunneling rate 
$\tau_{{\bf r},{\bf r}'}=\tau_1\ll\tau_0$. A corresponding complex mean-field
state is
\beq
|\Psi_D\rangle=\prod_{<{\bf r},{\bf r}'>\in D}
{1\over\sqrt{2}}(e^{i\varphi_{\bf r}}A_{\bf r}^\dagger
+e^{i\varphi_{{\bf r}'}}A_{{\bf r}'}^\dagger)|0\rangle,
\label{dimer}
\eeq
where $D$ is a set of dimers $\{<{\bf r},{\bf r}'>\}$ with 
$\tau_{{\bf r},{\bf r}'}=\tau_0$. This state is a lattice 
generalization of the two-site state of Eq.(\ref{gstate}). The Hamiltonian
of the two-component Bose gas at single-filling
(\ref{hcham}) reads in this mean-field approximation
\beq
-\langle \Psi_D|P_0H_1^2P_0|\Psi_D\rangle
=-{\tau_0^2\over2}t_\uparrow t_\downarrow\sum_{<{\bf r},{\bf r}'>\in D}
\cos(\varphi_{\bf r}-\varphi_{{\bf r}'})
-{t_\uparrow^2+t_\downarrow^2\over 4}\sum_{<{\bf r},{\bf r}'>}
\tau_{{\bf r},{\bf r}'}^2.
\label{ham3}
\eeq
For $t_\uparrow t_\downarrow=0$ this result agrees with the Hamiltonian
of solution (\ref{sol2}) but has a lower energy for any
$t_\uparrow t_\downarrow>0$. Moreover, the state $|\Psi_D\rangle$ has 
always a lower 
energy than the homogeneous mean-field state $|\Psi_{MF}\rangle_h$. This can
be summarized by comparing the ground-state energies: 
The difference between the ground-state energy of the homogeneous ($E_h$) 
state $|\Psi_{MF}\rangle_h$
and the inhomogeneous ($E_i$) state $|\Psi_{MF}\rangle_i$ is  
\[
E_h-E_i={(t_\uparrow-t_\downarrow)^2\over8}\sum_{<{\bf r},{\bf r}'>}
\tau_{{\bf r},{\bf r}'}^2
\]
and between the inhomogeneous state $|\Psi_{MF}\rangle_i$ and the dimerized
state $|\Psi_D\rangle$ is
\[
E_i-E_D={t_\uparrow t_\downarrow\over2}\sum_{<{\bf r},{\bf r}'>\in D}
\tau_{{\bf r},{\bf r}'}^2.
\]
It should be noticed that the homogeneous state $|\Psi_{MF}\rangle_h$ and
the dimerized state $|\Psi_D\rangle$ have the same ground-state energy
in the case of a single-component hard-core Bose gas, provided that
the dimers fill half of the lattice. This means that
the difference of the ground-state energies in the Hamiltonian of Eq.
(\ref{hcham}) is due to the repulsive interaction.

The Hamiltonian of Eq. (\ref{ham3}) creates short-range correlated 
fluctuations, since the set $D$ contains only isolated dimers. Therefore, 
it represents a Mott insulator.
If the modulation of the lattice is weak (i.e. $\tau_0\approx \tau_1$),
a summation of the state in Eq. (\ref{dimer}) over different dimer 
configurations is required. This may lead again to long-range correlated 
phase fluctuations, since the global $U(1)$ symmetry of the phase
fluctuations can be spontaneously broken.
Thus a phase transition from a Mott insulator at strong modulation to a
state with long-range correlated phase fluctuations at weak modulation is 
expected.

\section{Summary}

A two-component Bose gas with creation operators 
$a^\dagger_\uparrow$, $a^\dagger_\downarrow$ in an optical lattice with
lattice modulations and one particle per site is studied. 
By allowing only virtual tunneling to 
empty and doubly-occupied sites, an effective Hamiltonian is derived for
hard-core bosons, defined by the creation operator 
\[
A^\dagger = P_0a^\dagger_{{\bf r},\uparrow}a_{{\bf r},\downarrow}P_0,
\]
where $P_0$ is the projector on one-particle states. The effective Hamiltonian
decribes tunneling and a repulsive nearest-neighbor interaction between 
hard-core bosons. It is studied in terms of two types of mean-field states: 
a product of single-particle states and a dimerized state.
The ground-state energies of a homogeneous single-particle product state 
($E_h$), of a inhomogeneous single-particle product state ($E_i$), and of a 
dimerized state ($E_D$) are related as
\[
E_D\le E_i\le E_h,
\]
where the first equality sign holds for $t_\uparrow t_\downarrow=0$ and the
second for $t_\uparrow =t_\downarrow$.

The repulsive nearest-neighbor interaction prefers the
dimerized state, indicating a Mott insulator at least in the presence of 
a lattice modulation. This state is characterized by short-range
correlated phase fluctuations. For small or even vanishing modulation,
however, a superposition of different dimerized states may lead to a 
spontaneously broken $U(1)$ symmetry of the phase fluctuations. This would be 
accompanied by a Goldstone mode with long-range correlated phase fluctuations,
indicating the destruction of the Mott insulator and the creation of an 
ordered state in terms of the two components of the Bose gas. 

A similar model with $N$ components and hard-core interaction was studied in 
the $N\to\infty$ limit \cite{ziegler03}. It has a Mott-insulating phase
with ${\bar n}=1$ and indicates a symmetry-breaking phase for 
$0<{\bar n}<1$.

\vskip0.5cm

\noindent
This work was supported by the Deutsche Forschungsgemeinschaft through
Sonderforschungsbereich 484.

\section*{Appendix A}

Given is a sequence of projectors $P_j$ ($j\ge0$), defined by 
\[
P_{2k+1}=P_{2k-1}-P_{2k}\ \ \ (k\ge0)
\]
with initial condition $P_{-1}={\bf 1}$ and with the Hamiltonian $H$
through the properties
\beq
P_{2k}HP_{2k+1}=P_{2k}HP_{2k+2},\ \ \
P_{2k+1}HP_{2k}=P_{2k+2}HP_{2k}.
\label{recurse}
\eeq
With these projectors the identity Eq. (\ref{pidentity}) can be iterated.
The first step is to replace $P_0H P_1 (z-iH)_1^{-1}P_1HP_0$ with the 
right-hand side of the identity 
\[
P_0H P_1 (z-iH)_1^{-1}P_1HP_0
=P_0H P_2 (z-iH)_1^{-1}P_2HP_0
\]
such that Eq. (\ref{pidentity}) reads
\[
P_0(z-iH)^{-1}P_0=
(P_0(z-iH)P_0 + P_0H P_2 (z-iH)_1^{-1}P_2 HP_0)_0^{-1}.
\]
Now the expression $P_2 (z-iH)_1^{-1}P_2$ on the right-hand side can be 
rewritten by applying again Eq. (\ref{pidentity}) as
\[
P_2(z-iH)_1^{-1}P_2=
(P_2(z-iH)P_2 + P_2H P_3 (z-iH)_3^{-1}P_3 HP_2)_2^{-1}
\]
with $P_3=P_1-P_2$. Moreover, application of (\ref{recurse}) to the right-hand
side yields
\[
P_2(z-iH)_1^{-1}P_2=
(P_2(z-iH)P_2 + P_2H P_4 (z-iH)_3^{-1}P_4 HP_2)_2^{-1}.
\]
Iteration of this procedure leads to the recurrence relation
\beq
P_{2k}(z-iH)_{2k-1}^{-1}P_{2k}=
(P_{2k}(z-iH)P_{2k} + P_{2k}H P_{2k+2}(z-iH)_{2k+1}^{-1}P_{2k+2} HP_{2k}
)_{2k}^{-1}.
\label{recurr}
\eeq
Together with Eq. (\ref{pidentity}) this gives a continued-fraction 
representation of $P_0(z-iH)^{-1}P_0$.

\section*{Appendix B}

To evaluate the integral (\ref{int2}) it is convenient to define
\[
E_1={2\over N}\Delta E+E(1),\ \ \ \Delta E=[E(0)+E(2)]/2-E(1).
\]
Moreover, the spectral representation of $P_0H_1^2P_0$ is used
with the eigenvalues $\lambda_j$. This leads to 
\[
Z=\sum_j I(\lambda_j)
\]
with the integral
\[
I(\lambda)={1\over 2\pi i}\int_{-\infty}^\infty{e^{i\beta z}(z-iNE_1)\over
(z-iNE(1))(z-iNE_1)+\lambda}dz.
\]
The poles of the integrand are
\[
z_{\pm}={i\over 2}\Big[
N(E(1)+E_1)\pm\sqrt{N^2(E_1-E(1))^2+4\lambda}
\Big]
\]
such that the integral itself is
\[
I(\lambda)={e^{i\beta z_+}(z_+-iNE_1)-e^{i\beta z_-}(z_--iNE_1)\over
z_+-z_-}.
\]
The second term of the numerator dominates at large values $\beta$:
\beq
I(\lambda)\sim -{e^{i\beta z_-}(z_--iNE_1)\over z_+-z_-}.
\label{int0}
\eeq
Since
\[
z_+-z_-=i\sqrt{N^2(E_1-E(1))^2+4\lambda},
\]
\[
z_--iNE_1={i\over 2}\Big[
N(E(1)-E_1)-\sqrt{N^2(E_1-E_0)^2+4\lambda}
\Big]
\]
and 
\[
E_1-E(1)={2\over N}\Delta E,
\]
the expression in Eq. (\ref{int0}) reads
\[
I(\lambda)\sim {1\over2}
e^{-\beta(NE(1)+\Delta E)}e^{\beta\sqrt{(\Delta E)^2+\lambda}}
\Big[1+{\Delta E\over \sqrt{(\Delta E)^2+\lambda}}
\Big].
\]

\section*{Appendix C}

It is convenient to split the summation $\sum_{\sigma,\sigma'}$ 
in Eq. (\ref{pham}) into a diagonal part and an off-diagonal part:
\[
-P_0H_1^2P_0=
\]
\[
-{1\over 2}\sum_{<{\bf r},{\bf r}'>}\tau_{{\bf r},{\bf r}'}^2
\Big[\sum_{\sigma=\downarrow,\uparrow}
t_\sigma^2
P_0a^\dagger_{{\bf r},\sigma} a_{{\bf r},\sigma}a_{{\bf r}',\sigma} 
a^\dagger_{{\bf r}',\sigma}P_0
+2t_\uparrow t_\downarrow
P_0a^\dagger_{{\bf r},\downarrow} a_{{\bf r},\uparrow}
a^\dagger_{{\bf r}',\uparrow} a_{{\bf r}',\downarrow}P_0\Big].
\]
The projection $P_0$ acts individually at each lattice site ${\bf r}$, 
i.e., for ${\bf r}'\ne {\bf r}$ one can write
\[
P_0a^\dagger_{{\bf r},\sigma} a_{{\bf r},\sigma}
a_{{\bf r}',\sigma} a^\dagger_{{\bf r}',\sigma}P_0
=P_0a^\dagger_{{\bf r},\sigma} a_{{\bf r},\sigma}P_0
P_0a_{{\bf r}',\sigma} a^\dagger_{{\bf r}',\sigma}P_0
\]
and
\[
P_0a^\dagger_{{\bf r},\sigma} a_{{\bf r},\sigma'}
a^\dagger_{{\bf r}',\sigma'} a_{{\bf r}',\sigma}P_0
=P_0a^\dagger_{{\bf r},\sigma} a_{{\bf r},\sigma'}P_0
P_0a^\dagger_{{\bf r}',\sigma'} a_{{\bf r}',\sigma}P_0.
\]
With this it is possible to define operators on the $P_0$-projected 
Hilbert space (i.e. the space with one particle per site) as
\[
A^\dagger_{{\bf r}}=P_0a^\dagger_{{\bf r},\uparrow}a_{{\bf r},
\downarrow}P_0,\ \ 
A_{{\bf r}}=P_0a^\dagger_{{\bf r},\downarrow}a_{{\bf r},\uparrow}P_0.
\]
When $|\downarrow\rangle$ is formally identified with a vacuum state and 
$|\uparrow\rangle$ with a particle,
$A^\dagger$ ($A$) is a creation (annihilation) operator for a hard-core
boson on the $P_0$-projected Hilbert space. Moreover, it is
\begin{equation}
P_0a_{{\bf r},\sigma}a^\dagger_{{\bf r},\sigma}P_0=P_0
-P_0a^\dagger_{{\bf r},\sigma}a_{{\bf r},\sigma}P_0,
\label{sum1}
\end{equation}
and the operators satisfy the identities
\begin{equation}
A^\dagger_{{\bf r}}A_{{\bf r}}
=P_0a^\dagger_{{\bf r},\uparrow}a_{{\bf r},\downarrow}P_0
a^\dagger_{{\bf r},\downarrow}a_{{\bf r},\uparrow}P_0
=P_0a^\dagger_{{\bf r},\uparrow}a_{{\bf r},\uparrow}P_0
=P_0a_{{\bf r},\downarrow}a^\dagger_{{\bf r},\downarrow}P_0
\label{rel0}
\end{equation}
\begin{equation}
A_{{\bf r}}A^\dagger_{{\bf r}}
=P_0a^\dagger_{{\bf r},\downarrow}a_{{\bf r},\uparrow}
P_0a^\dagger_{{\bf r},\uparrow}a_{{\bf r},\downarrow}P_0
=P_0a^\dagger_{{\bf r},\downarrow}a_{{\bf r},\downarrow}P_0
=P_0a_{{\bf r},\uparrow}a^\dagger_{{\bf r},\uparrow}P_0.
\end{equation}
Thus $A^\dagger_{{\bf r}}A_{{\bf r}}$ is the particle number operator for the
hard-core bosons.
With (\ref{sum1}) and (\ref{rel0}) the Hamiltonian $-P_0H_1^2P_0$
can be written in terms of the hard-core Bose operators as
\[
-P_0H_1^2P_0
=-\sum_{<{\bf r},{\bf r}'>}\tau_{{\bf r},{\bf r}'}^2\Big[
t_\uparrow t_\downarrow A^\dagger_{{\bf r}}A_{{\bf r}'}
+{t_\uparrow^2+t_\downarrow^2\over2}({\bf 1}-A^\dagger_{{\bf r}}
A_{{\bf r}})A^\dagger_{{\bf r}'}A_{{\bf r}'}
\Big].
\]

\end{document}